\documentclass[acmsmall,screen,nonacm]{acmart}

\setcopyright{none}

\usepackage{algorithmic}
\usepackage{textcomp}
\usepackage{xspace}
\usepackage{multirow}
\usepackage{makecell}
\usepackage{tabularx}
\usepackage{enumitem}
\usepackage{placeins}
\usepackage{makecell}

\usepackage[table]{xcolor}
\usepackage{colortbl}

\newcommand{\fixme}[1]{\textcolor{orange}{{\it [fixme]}}}

\newcommand{\phead}[1]{\noindent {\bf #1}}

\newcommand{\results}[1]{\noindent {\textbf{\textit{#1}}}}
\newcommand{\findings}[1]{\noindent {\textbf{Findings.} \textbf{\textit{#1}}}}
\newcommand{\summarybox}[2]{%
  \par\vskip\smallskipamount\noindent
  \begingroup
  \setlength{\fboxsep}{6pt}%
  \colorbox{gray!10}{%
    \parbox{\dimexpr\linewidth-2\fboxsep\relax}{%
      \textbf{#1} #2%
    }%
  }%
  \endgroup\par
}
\newcommand{\rqanswer}[2]{\summarybox{Answer to RQ#1:}{#2}}
\newcommand{\observation}[1]{\summarybox{Observation:}{#1}}

\AtBeginDocument{%
  }

\begin{document}

\title{From Finding to Fixing: How Useful is Fault Localization for Agentic Program Repair?}

\author{Xinyu Shi}
\affiliation{%
  \institution{University of Alberta}
  \city{Edmonton }
  \country{Canada}}
\email{xshi12@ualberta.ca}

\author{Zhou Yang}
\affiliation{%
  \institution{University of Alberta}
  \city{Edmonton }
  \country{Canada}}
\email{zy25@ualberta.ca}

\author{An Ran Chen}
\affiliation{%
  \institution{University of Alberta}
  \city{Edmonton }
  \country{Canada}}
\email{anran6@ualberta.ca}

\renewcommand{\shortauthors}{Shi et al.}

\begin{abstract}
Fault localization (FL) aims to identify code locations responsible for software failures and is commonly evaluated by how accurately it ranks known fault locations.
Large language model (LLM) agents are changing program repair and are increasingly used to search and reason over repositories.
Although recent LLM-based FL methods have improved localization accuracy, their impact on agentic repair remains unclear.
To address this gap, we conduct a systematic empirical study of FL in agentic repair.
We evaluate six recent FL methods on SWE-bench Verified and Defects4J with three LLMs under a common repair framework and analyze 16,248 repair trajectories.
We also examine gold locations and bug-cause descriptions as forms of FL guidance across different repair contexts.
The studied FL methods do not consistently improve repair success, and higher localization accuracy does not consistently lead to larger repair gains.
Yet FL guides agents' early search and helps them reach relevant code sooner.
Accurate locations can reduce repair token use by up to 39.2\%, but current FL methods often consume more tokens than they save during repair.
Experiments with gold guidance suggest that FL has greater potential when bug inputs contain fewer location clues.
Providing bug causes can further improve repair success beyond locations alone.
We also find that ground truth from a single reference patch does not fully capture the locations used by successful repairs.
These findings help clarify the value of FL in agentic repair and inform future work on its design and evaluation.
\end{abstract}

\ccsdesc[500]{Software and its engineering~Software testing and debugging}

\keywords{fault localization, program repair, large language models}


\maketitle

\section{Introduction}
\label{Introduction}

Debugging software failures requires developers to understand what went wrong, locate relevant code, and determine a fix.
Fault localization (FL) aims to identify code locations that may be responsible for a software failure.
Given bug information such as failing tests or an issue description, FL methods rank candidate code locations by their likelihood of containing the fault.
Earlier approaches~\cite{jones2005empirical,abreu2007accuracy, zhou2012should} used program execution information, such as test coverage, or information retrieval between bug reports and source code.
Learning-based methods~\cite{sohn2017fluccs,lou2021boosting, li2019deepfl} used features of the code and its behavior to learn how to rank candidate locations.
More recent methods~\cite{kang2024quantitative,10891926,chen2025locagent,jiang2025issue} use large language models (LLMs) to interpret bug information, reason about program behavior, and identify suspicious code locations.
Despite these methodological changes, FL is still commonly formulated as a retrieval and ranking problem: methods return suspicious files, methods, or lines, and are typically evaluated by how highly known fault locations are ranked, using metrics such as Top-$N$~\cite{kochhar2016practitioners,kang2024quantitative,10891926}.

Meanwhile, LLMs and agents are changing software engineering workflows.
LLM agents combine language models with tools and iterative interaction, allowing them to take actions, observe the results, and adjust their subsequent decisions~\cite{yao2023react}.
In program repair, such agents can tackle repository-level tasks end to end~\cite{wang2025openhands,tang2024codeagent}.
Starting from a bug report or failing test, an agent can locate relevant code, modify the implementation, and validate the resulting patch through testing~\cite{yang2024swe,bouzenia2025repairagent}.
As agents are applied to an expanding range of software engineering tasks, they are increasingly becoming a common part of developers' coding workflows.

However, the value of FL for agentic repair remains unclear, especially as agents can search for relevant code and make repairs within an end-to-end workflow.
Existing FL studies still focus primarily on localization accuracy, with limited evaluation of how FL affects downstream agentic repair.
Beyond repair outcomes, existing studies also rarely examine how FL guidance affects agent behavior during repair.
More broadly, as agents take a more active role in repository exploration and repair, the role and value of FL in agentic repair deserve closer examination.

To address these gaps, we investigate FL in agentic repair from three perspectives: whether FL improves repair success, how its guidance affects the repair process, and when it provides additional value. We organize the study around three research questions:
\begin{itemize}[]
    \item RQ1: Does FL improve agentic repair?
    \item RQ2: How does FL affect the agent's repair process?
    \item RQ3: When does FL provide additional value for agentic repair?
\end{itemize}

Our study covers two representative FL settings. We reproduce six recent FL methods on SWE-bench Verified and Defects4J v3.0.1, covering issue-based and test-based localization, and evaluate them with three repair models under the same agentic repair framework. 
For each benchmark and repair model, we evaluate repair without FL and with each of the three corresponding FL methods. We first compare repair with and without FL guidance to determine whether localization improves repair success and whether stronger localization translates into larger repair gains. 
We then analyze 16,248 distinct repair trajectories across these settings to measure changes in repository navigation, use of localization guidance, responses to misleading locations, and resource consumption.
We also complement this analysis with a manual study.
Finally, we move beyond the common contexts in which FL is typically evaluated and examine its role under broader agentic repair settings. We study cases with fewer location clues in the issue, multimodal bug inputs, and larger, more distributed repairs, and further examine whether conventional reference locations adequately represent where successful agents actually make repairs.

Our results show a more nuanced relationship between FL and agentic repair than localization accuracy alone suggests.
The studied FL methods do not consistently improve repair success.
Across the 18 FL-repair settings, adding FL changes pass@1 by \(-4.45\) to \(+0.40\) percentage points relative to No-FL, with decreases in 17 settings. Higher localization accuracy does not consistently translate into larger repair gains.
Even when FL guidance is completely misleading, agents can reassess it and often still reach relevant code.
Nevertheless, FL strongly shapes agents' initial search and helps them reach relevant code earlier.
Accurate locations reduce repair token use by up to 39.2\%, but the overhead of existing FL methods often outweighs downstream token savings.
Our experiments suggest that FL has greater potential when bug inputs contain fewer direct location clues and that bug-cause information can provide further repair gains beyond locations alone.
Conventional FL ground truth does not fully capture the locations used by successful repairs.
Together, these findings clarify the value and limitations of FL in agentic repair and provide implications for future research on its design and evaluation.
In summary, we make the following contributions:
\begin{itemize}[leftmargin=*]
    \item \textbf{Downstream evaluation of FL for agentic repair.} We systematically evaluate six recent FL methods on two benchmarks with three backbone models under a common agentic repair framework.
    Our results show that these methods do not consistently improve repair success, and higher localization accuracy does not consistently translate into larger repair gains.
    These findings highlight the need to evaluate FL through its contribution to downstream repair.

    \item \textbf{Understanding how FL affects agentic repair.} Through large-scale trajectory analysis and manual inspection, we show that FL primarily shapes agents' early repository search and helps them reach relevant code sooner.
    Agents assess the suggested locations through further exploration and can move beyond misleading guidance.
    FL often reduces repair effort, but localization overhead frequently outweighs the downstream token savings.

    \item \textbf{Rethinking the value of FL for agentic repair.} Experiments with gold guidance show that accurate locations can reduce repair token use, that localization provides greater value when bug inputs contain fewer location clues, and that bug-cause information can improve repair success beyond locations alone.
    We also find that conventional FL ground truth does not fully capture the locations used by successful repairs.
    These findings motivate on-demand localization, guidance that combines locations with diagnostic information, and evaluation that accounts for alternative repair locations.

\end{itemize}

\section{Background \& Related Work}
\label{sec:background-related-work}






\subsection{Fault Localization}

\begin{table*}[t]
\centering
\caption{Overview of recent LLM-based FL methods.}
\label{tab:fl_methods}
\small
\setlength{\tabcolsep}{2.5pt}
\renewcommand{\arraystretch}{1.0}
\renewcommand{\tabularxcolumn}[1]{m{#1}}
\begin{tabularx}{\textwidth}{@{}l
  >{\raggedright\arraybackslash}m{0.215\textwidth}
  >{\raggedright\arraybackslash}X
  l
  c@{}}
\toprule
\textbf{Method} &
\textbf{Bug input} &
\textbf{Benchmark} &
\textbf{Output} &
\textbf{Agentic repair eval.} \\
\midrule

AutoFL~\cite{kang2024quantitative}
& Failing test
& Defects4J v1.0, BugsInPy
& Method
& \textcolor{red!75!black}{$\times$} \\

\rowcolor{black!6}
LLM4FL~\cite{rafi2024multi}
& Failing test
& Defects4J v2.0.0
& Method
& \textcolor{red!75!black}{$\times$} \\

SoapFL~\cite{10891926}
& Failing test
& Defects4J v1.2.0
& Method
& \textcolor{red!75!black}{$\times$} \\

\rowcolor{black!6}
FlexFL~\cite{xu2025flexfl}
& Bug report/failing test
& Defects4J v2.0.0, GHRB
& Method
& \textcolor{red!75!black}{$\times$} \\

MemFL~\cite{yeo2025improving}
& Failing test
& Defects4J v1
& Method
& \textcolor{red!75!black}{$\times$} \\

\rowcolor{black!6}
FaR-Loc~\cite{shi2025enhancing}
& Failing test
& Defects4J v1.2.0/v2.0.0
& Method
& \textcolor{red!75!black}{$\times$} \\

\midrule

LocAgent~\cite{chen2025locagent}
& Issue description
& SWE-bench Lite, LocBench
& Line
& \textcolor{red!75!black}{$\times$} \\

\rowcolor{black!6}
CoSIL~\cite{jiang2025issue}
& Issue description
& SWE-bench Lite/Verified
& Function
& \textcolor{red!75!black}{$\times$} \\

OrcaLoca~\cite{3780338.3783277}
& Issue description
& SWE-bench Lite
& Function
& \textcolor{red!75!black}{$\times$} \\

\rowcolor{black!6}
SHERLOC~\cite{tamoyan2026sherloc}
& Issue description
& SWE-bench Lite/Verified
& Line
& \textcolor{green!50!black}{$\checkmark$} \\

SWERank~\cite{gangi2026swerank}
& Issue description
& SWE-bench Lite, LocBench
& Function
& \textcolor{red!75!black}{$\times$} \\

\rowcolor{black!6}
GraphLocator~\cite{liu2026graphlocator}
& Issue description
& SWE-bench Lite, LocBench, Multi-SWE-bench (Java)
& Function
& \textcolor{green!50!black}{$\checkmark$} \\

\bottomrule
\end{tabularx}
\end{table*}

\noindent FL is a fundamental but time-consuming task in software debugging, with the goal of identifying faulty code elements. Research indicates that more than half of the debugging time is spent on finding the fault location \cite{bohme2017bug, alaboudi2021exploratory,chen2021demystifying}. 
To reduce manual effort, researchers have introduced various automated FL techniques.
FL methods have been studied under different input settings. Some start with a file already known to be buggy, while others rely on a triggering test and its failure output, a bug report, or a GitHub issue.
In each setting, the method uses the available information to identify suspicious code locations.
Early spectrum-based methods used passing and failing test coverage to score suspicious code locations~\cite{jones2005empirical}.
Learning-based methods later used features of the code and its behavior to learn how to rank candidates.
Recent methods use large language models (LLMs) to interpret bug information and explore repositories~\cite{kang2024quantitative,chen2025locagent}.
Despite these changes, the task is still commonly framed as retrieving candidate locations and ranking them by how likely they are to contain the fault.
Recent methods report gains in localization accuracy, but their evaluations still focus mainly on whether those locations match reference fault locations~\cite{kang2024quantitative,shi2025enhancing}.

\phead{FL Methods.}
We reviewed representative recent localization methods to examine how they formulate the task.
Table~\ref{tab:fl_methods} summarizes their bug inputs, evaluation benchmarks, output granularity, and whether their cited papers evaluate agentic repair.
The methods in the table mainly use two forms of bug input: issue descriptions and triggering-test information.

\noindent\textit{(1) Issue-based localization.}
These methods search a repository from a natural-language issue description.
LocAgent~\cite{chen2025locagent} represents code entities and their dependencies as a heterogeneous graph, which an LLM agent traverses to find relevant code.
CoSIL~\cite{jiang2025issue} narrows the search in two stages: it first explores a module call graph, then searches a function call graph while pruning unrelated paths and context.
OrcaLoca~\cite{3780338.3783277} instead controls the agent's search through priority-based action scheduling, action decomposition, and distance-aware context pruning.
GraphLocator~\cite{liu2026graphlocator} addresses the gap between a reported symptom and its underlying cause by building a causal issue graph that links sub-issues to code entities.
SHERLOC~\cite{tamoyan2026sherloc} pairs repository tools with self-recovery mechanisms and produces diagnostic findings alongside candidate locations.
SWERank~\cite{gangi2026swerank} combines a code retriever with an LLM reranker, both fine-tuned on issue descriptions paired with relevant functions identified from code changes.
Together, these methods show different ways to search from an issue, from graph-guided exploration to learned code ranking.

\noindent\textit{(2) Test-based localization.}
These methods use a failing test and information such as its stack trace or code coverage to guide the search.
AutoFL~\cite{kang2024quantitative} provides a set of exploration functions that let an LLM retrieve information about classes, methods, and comments within code covered by failing tests to pinpoint the fault location.
SoapFL~\cite{10891926} follows a hierarchical approach that progressively narrows the search to suspicious files and then to methods within those files.
It uses LLM-generated code summaries to guide localization at each level.
FlexFL~\cite{xu2025flexfl} first gathers candidate methods using several conventional localization techniques and then asks an open-source LLM to inspect and rerank them; it can also use bug reports when available.
LLM4FL~\cite{rafi2024multi} starts from spectrum-based rankings, divides coverage information into manageable groups for multiple agents, and uses self-reflection to revise the final ranking.
MemFL~\cite{yeo2025improving} adds external memory containing project summaries and insights from earlier localization attempts.
FaR-Loc~\cite{shi2025enhancing} turns the failed test into a natural-language description of the failing behavior, retrieves semantically related covered methods, and uses an LLM to rerank them.
These methods mainly target method-level localization, but differ in how they narrow the candidates and use failure context.

\subsection{Agentic Program Repair}

\phead{Program repair methods.}
Automated program repair (APR) aims to generate patches that fix software bugs. Early APR research explored how tests and program analysis could guide the search for useful code changes. GenProg~\cite{weimer2009automatically} used test results to guide a genetic search over program edits. SemFix~\cite{nguyen2013semfix} first ranked suspicious statements, then used symbolic execution and program synthesis to construct a repair. In both systems, finding where to edit was part of producing the patch. As LLMs became available for code generation, some repair studies used them to generate patches for a supplied buggy file or code span. Repilot~\cite{wei2023copiloting} combines an LLM with a code-completion engine to avoid invalid patch candidates. ChatRepair~\cite{xia2024automated} uses feedback from failing tests to revise patches through a conversation with the LLM.

LLM-based repair later expanded to repository-level problems described in GitHub issues. These tasks require a method to find relevant code as well as generate a patch. Some methods kept these steps in a fixed pipeline. Agentless~\cite{xia2024agentless}, for example, separates localization, repair, and patch validation into three phases. The LLM contributes within each phase, while the order of the phases and the handoff between them are predefined.

More recent agentic approaches give the LLM a larger role in directing the repair process. An agent can choose tool actions, inspect their results, and decide what to do next. Repair-focused agents use this flexibility in different ways. RepairAgent~\cite{bouzenia2025repairagent} can move between gathering information, patching, and validation as feedback arrives. AutoCodeRover~\cite{zhang2024autocoderover} uses iterative search over program structure to find relevant code before producing a patch. General software agents also support repository-level repair: SWE-agent~\cite{yang2024swe} provides an interface for searching, viewing, and editing repository code; OpenHands~\cite{wang2025openhands} lets agents write code, use a command line, and browse the web; and mini-SWE-agent~\cite{minisweagent2026docs} uses a small set of shell-based actions. In these systems, finding the bug location becomes a natural part of the repair process.

\phead{Localization and program repair.}
While agents can search for faulty code on their own, it is unclear whether supplying them with FL results improves repair.
Of the twelve methods in Table~\ref{tab:fl_methods}, six report no downstream repair evaluation in their cited papers, four evaluate repair only through fixed pipelines, and two include agentic repair evaluation.
LocAgent~\cite{chen2025locagent}, CoSIL~\cite{jiang2025issue}, and OrcaLoca~\cite{3780338.3783277} pass their locations to the Agentless pipeline, while SWERank~\cite{gangi2026swerank} uses the SWE-Fixer pipeline.
These evaluations do not show how a repair agent uses the guidance during repository exploration.
GraphLocator~\cite{liu2026graphlocator} evaluates both Agentless and Trae Agent, while SHERLOC~\cite{tamoyan2026sherloc} provides its locations and diagnostic findings to SWE-agent and OpenHands.
GraphLocator also evaluates LocAgent, CoSIL, and SWERank as localization baselines within Trae Agent.
Thus, agentic repair evaluation remains concentrated in a small number of studies.
SHERLOC reports an average improvement in repair, but its guidance does not improve the resolve rate in every agent and model setting.
This variation calls for a closer study of when and how FL helps agentic repair.

Several studies directly examine how localization guidance affects repair, but most test fixed pipelines rather than agents that search the repository during repair. In these pipelines, an LLM generates a patch directly from selected code context. RGFL~\cite{sepidband2026rgfl} changes the localization stages in Agentless, but repair still follows its fixed pipeline. Townsend et al.~\cite{townsend2026study} vary the granularity of locations passed to a modified Agentless pipeline. Because those locations come from ground-truth patches, this setup does not capture errors made by real FL methods. Sepidband et al.~\cite{sepidband2026role} examine the code context given to an LLM, but the model generates patches from a prompt without tools or further repository exploration. Loc2Repair~\cite{awad2026loc2repairframeworkevaluatingimpact} takes a step toward agentic repair by testing file-level hints for GitHub issues, but does not examine method-level or test-based guidance. What remains unclear is how FL methods with different bug inputs affect an agent's search, patch choices, and response to misleading locations.

\observation{Based on their bug inputs, recent FL methods mainly fall into two categories: issue-based and test-based. The impact of FL on agentic repair still needs systematic study.}

\section{Experimental Setup}
\label{sec:experimental-setup}

\begin{table*}[t]
\centering
\caption{Overview of the study settings.}
\label{tab:rq1-settings}
\footnotesize
\setlength{\tabcolsep}{4pt}
\renewcommand{\arraystretch}{1.05}
\begin{tabularx}{\textwidth}{@{}l
  >{\hsize=.75\hsize\linewidth=\hsize\raggedright\arraybackslash}X
  >{\hsize=1.25\hsize\linewidth=\hsize\raggedright\arraybackslash}X r@{}}
\toprule
\textbf{Benchmark} &
\textbf{Bug input} &
\textbf{FL guidance} &
\textbf{\# Settings} \\
\midrule
\rowcolor{black!5}
\multicolumn{4}{@{}l}{\textbf{RQ1--RQ2}\quad (GPT-5.6 Luna, DeepSeek-v4 Flash, Gemini 3.1 Flash Lite)} \\
\addlinespace[2pt]
SWE-bench Verified & 500 issues & No-FL, LocAgent, CoSIL, OrcaLoca & $4\times3=12$ \\
Defects4J v3.0.1 & 854 failing tests & No-FL, AutoFL, SoapFL, FlexFL & $4\times3=12$ \\
\midrule
\rowcolor{black!5}
\multicolumn{4}{@{}l}{\textbf{RQ3}\quad (GPT-5.6 Luna)} \\
\addlinespace[2pt]
\multirow[t]{2}{*}{SWE-bench Verified} & 500 issues & \multirow[c]{4}{=}{\raggedright\textit{For all RQ3 settings:}\\No-FL, gold file, gold symbol,\\bug cause, gold symbol + bug cause} & 5 \\
 & 500 LR-issues & & 5 \\
SWE-bench Multimodal & 480 issues + images & & 5 \\
SWE-bench Pro (bug subset) & 221 issues & & 5 \\
\midrule
\multicolumn{3}{@{}l}{\textbf{Total}} & \textbf{44} \\
\bottomrule
\end{tabularx}
\end{table*}

Our study has two stages: we first examine how current FL methods affect repair outcomes and agent behavior (RQ1--RQ2), then investigate when FL provides additional value across repair contexts and forms of guidance (RQ3).
Table~\ref{tab:rq1-settings} summarizes our experimental settings across RQ1--RQ3.
For RQ1--RQ2, we evaluate two benchmark families that represent the two main bug input settings considered in recent FL research: issue descriptions in SWE-bench Verified and failing-test information in Defects4J v3.0.1.
For each benchmark, we select three FL methods and evaluate each method with all three repair models: GPT-5.6 Luna, DeepSeek-v4 Flash, and Gemini 3.1 Flash Lite.
This gives 18 With-FL settings and six No-FL baselines, for a total of 24 repair settings and 18 paired With-FL/No-FL comparisons.
RQ3 evaluates five guidance conditions in four repair contexts using GPT-5.6 Luna, giving 20 settings including No-FL.
We compare repair with and without FL, analyze the resulting repair trajectories and resource use, and conduct targeted experiments to examine the effects of localization cues in issue inputs and the suitability of conventional FL ground truth for agentic repair.

\subsection{Benchmarks}
\label{sec:benchmarks}

We evaluate on two benchmark families, SWE-bench Verified~\cite{swebenchverifieddocs} and Defects4J v3.0.1~\cite{defects4jrepo}, which cover the two major FL input settings considered in this study: issue descriptions and triggering-test information. Both benchmarks are widely used in recent FL research~\cite{jiang2025issue,tamoyan2026sherloc,10891926,xu2025flexfl} and together cover issue-based and test-based localization settings.



\begin{itemize}[leftmargin=*]
    \item \textbf{SWE-bench Verified.} SWE-bench Verified contains 500 human-validated Python repository-level repair tasks. Each task provides a natural-language issue description and a repository snapshot. We use SWE-bench Verified rather than SWE-bench Lite because Verified covers a broader set of repository-level repairs, including tasks whose gold patches modify multiple files.


    \item \textbf{Defects4J v3.0.1.} Defects4J contains real-world Java bugs with buggy and fixed program versions and tests that expose the failures. We use all 854 bugs in version 3.0.1.

\end{itemize}

\subsection{Fault Localization Methods}
\label{sec:studied-methods}

We study six recently published LLM-based FL methods that have been evaluated on widely used FL benchmarks and represent different designs in current FL research.
We select three methods for each benchmark: LocAgent, OrcaLoca, and CoSIL for SWE-bench Verified, and AutoFL, SoapFL, and FlexFL for Defects4J.
These methods cover different localization strategies, including graph-based search, tool-based code exploration, hierarchical localization, and combinations of LLMs with non-LLM localization techniques.
We reproduce all six methods in our experimental setup and run each method with each of the selected LLMs for a consistent comparison.

\begin{itemize}[leftmargin=*]
\item \textbf{LocAgent.}
LocAgent~\cite{chen2025locagent} represents a repository as a graph of code entities and their relationships.
An LLM agent uses this graph to search for code relevant to an issue.

\item \textbf{OrcaLoca.}
OrcaLoca~\cite{3780338.3783277} uses an LLM agent for issue localization.
It combines priority-based action scheduling, action decomposition with relevance scoring, and context pruning to guide the search.

\item \textbf{CoSIL.}
CoSIL~\cite{jiang2025issue} performs issue localization through iterative code graph search.
It uses an LLM to explore repository context and narrow the search to relevant code entities.

\item \textbf{AutoFL.}
AutoFL~\cite{kang2024quantitative} uses an LLM in an iterative tool-calling loop to explore code covered by the failing tests.
The LLM uses tools such as \texttt{readFile} and \texttt{readMethod} to inspect relevant code and narrow down suspicious locations.

\item \textbf{SoapFL.}
SoapFL~\cite{10891926} performs localization in a hierarchical manner.
It first identifies suspicious files and then examines methods within those files.
It also uses LLM-generated summaries of the code to support localization decisions at different levels.

\item \textbf{FlexFL.}
FlexFL~\cite{xu2025flexfl} combines LLM-based reasoning with non-LLM localization techniques such as spectrum-based FL and retrieval techniques.
These techniques first help narrow the search space, after which an LLM inspects the candidate methods and refines the localization results.
\end{itemize}

\subsection{Repair Models and Agent}
\label{sec:selected-llms}

We use GPT-5.6 Luna, DeepSeek-v4 Flash, and Gemini 3.1 Flash Lite as our repair models.
These models were among the latest available at the time of our experiments.
They come from different companies and include both closed-source and open-weight models, reflecting different model choices for real-world agentic repair.
This choice allows us to examine whether the effects of FL vary across repair models.
All three models are designed to offer efficient inference, making them practical choices for repair experiments across our benchmarks and FL methods.

\begin{itemize}[leftmargin=*]
\item \textbf{GPT-5.6 Luna.}
GPT-5.6 Luna is a closed-source model from OpenAI~\cite{openai2026gpt56luna}.
It offers six reasoning effort settings, \texttt{none}, \texttt{low}, \texttt{medium}, \texttt{high}, \texttt{xhigh}, and \texttt{max}.
We use \texttt{medium} in our experiments.

\item \textbf{DeepSeek-v4 Flash.}
DeepSeek-v4 Flash is an open-weight model from DeepSeek~\cite{deepseek2026v4flash}.
Its weights and supporting code are available under the MIT license.
The official API supports \texttt{low}, \texttt{high}, and \texttt{max} reasoning effort~\cite{deepseek2026apiupdates}.
We use \texttt{high} in our experiments.

\item \textbf{Gemini 3.1 Flash Lite.}
Gemini 3.1 Flash Lite is a closed-source model from Google~\cite{google2026gemini31flashlite}.
It offers four thinking levels, \texttt{minimal}, \texttt{low}, \texttt{medium}, and \texttt{high}.
We use \texttt{medium} in our experiments.
\end{itemize}


We use mini-swe-agent 2.4.6~\cite{minisweagent2026docs} as the repair framework for all three LLMs.
The mini-swe-agent framework uses Bash as its only tool, allowing the agent to search the repository, read and edit source files, and run tests through shell commands.
This simple interface helps reduce bias from specialized tool designs and makes agent behavior easier to analyze.
At each step, the model is prompted to provide a \texttt{thought} field explaining its reasoning and tool calls to perform actions.
It then uses the command outputs to decide what to do next.
The recorded thoughts and actions allow us to trace the model's stated reasoning and its response to FL guidance.
For each case, we set a cost limit of \$3 and a step limit of 250 to reduce the risk that repair outcomes are affected by early termination due to resource limits.

\section{Results and Findings}

\subsection{RQ1: Does FL improve agentic repair?}
\label{sec:RQ1}

RQ1 examines whether FL improves repair success in agentic repair and whether its effect depends on localization quality.
As described in Sections~\ref{sec:benchmarks}--\ref{sec:selected-llms}, we study two benchmarks, three FL methods for each benchmark, and three repair models, giving 18 With-FL settings and six No-FL baselines.
For each benchmark and repair model, we first run mini-swe-agent without FL input and then run it with the output of each FL method added to the task description as extra guidance.
The same No-FL baseline is shared across the three FL methods for each benchmark and repair model.
To make the comparison fair, the two runs differ only in whether FL guidance is provided.
The agent still has access to the full repository and the same repair tools, and it can inspect or edit code beyond the suggested locations.
Across these 24 repair settings, the experiments produce 16,248 distinct repair trajectories, including 6,000 for SWE-bench Verified and 10,248 for Defects4J v3.0.1.

\begin{figure}
\centering
\includegraphics[width=\linewidth]{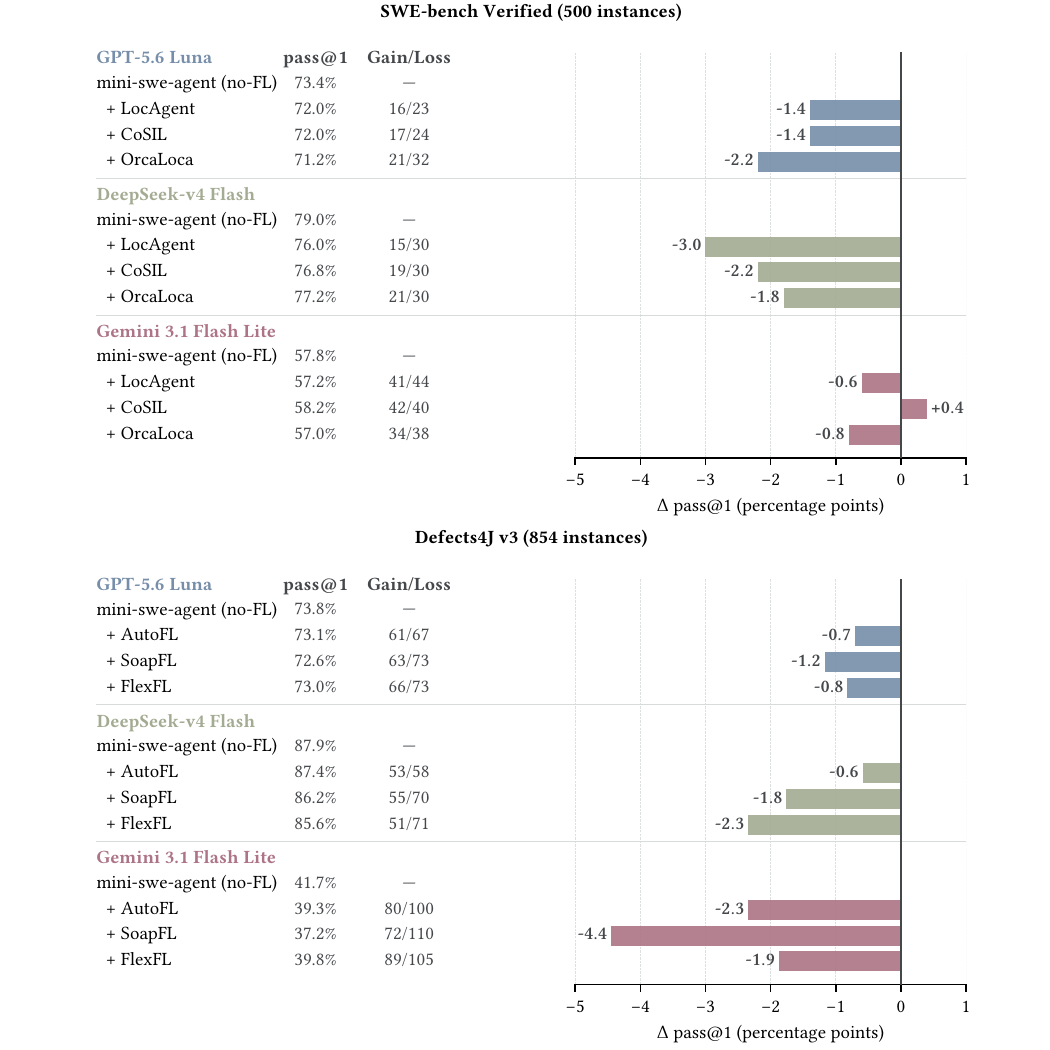}
\caption{Overall paired repair effects across the 18 With-FL settings.}
\Description{Paired changes in repair success and the numbers of targets helped or harmed by adding FL across the 18 With-FL settings.}
\label{fig:rq1-1-overall-effect}
\end{figure}

\subsubsection{RQ1.1: Does FL Improve Repair Success?}
\label{sec:RQ1-success}
\leavevmode\par

\phead{Design.}
For each of the 18 With-FL settings, we compare the repair outcome of every bug with FL against the corresponding outcome without FL.
We preserve each method's complete list of locations and its original format, and present the list as suspicious locations in the repair prompt.
For both benchmarks, a bug is considered successfully repaired if the generated patch passes the test suite used by the corresponding benchmark's official evaluator.
We measure repair success using pass@1, the percentage of bugs successfully repaired with one attempt.
We report pass@1 with and without FL and their difference in percentage points.
To further show how FL changes repair outcomes for individual bugs, we also count gains and losses.
A gain occurs when a bug is repaired with FL but not without it, while a loss occurs when the opposite happens.
We use a two-sided exact McNemar test~\cite{mcnemar1947sampling} for each setting to compare the paired gains and losses, and adjust the 18 $p$-values using Holm's method~\cite{holm1979simple}.


\findings{The studied FL methods do not consistently improve repair success.}
As shown in Figure~\ref{fig:rq1-1-overall-effect}, the change in pass@1 across the 18 With-FL settings ranges from \(-4.45\) to \(+0.40\) percentage points and is negative in 17 settings.
The only positive change is a small increase of \(+0.40\) percentage points for Gemini 3.1 Flash Lite with CoSIL on SWE-bench Verified.
None of the six FL methods consistently improves repair success across the three repair models.
In 16 of the 18 With-FL settings, the unadjusted McNemar $p$-value is at least 0.05, indicating no significant imbalance between gains and losses.
The two exceptions both show decreases in repair success: DeepSeek-v4 Flash with LocAgent ($p=0.036$) and Gemini 3.1 Flash Lite with SoapFL ($p=0.006$).
After Holm adjustment across the 18 With-FL settings, neither remains below 0.05 (adjusted $p=0.607$ and $p=0.107$, respectively).

\results{Nevertheless, FL changes which individual bugs are repaired.}
Every setting contains both gains and losses.
For example, GPT-5.6 Luna with LocAgent produces 16 gains and 23 losses, while Gemini 3.1 Flash Lite with LocAgent produces 41 gains and 44 losses.
These changes show that a small overall difference in pass@1 does not mean that FL has little influence on repair outcomes.
This motivates us to examine how the repair benefit of FL varies with localization quality, even when its overall effect on repair success is small.

\begin{figure}
\centering
\includegraphics[width=\linewidth]{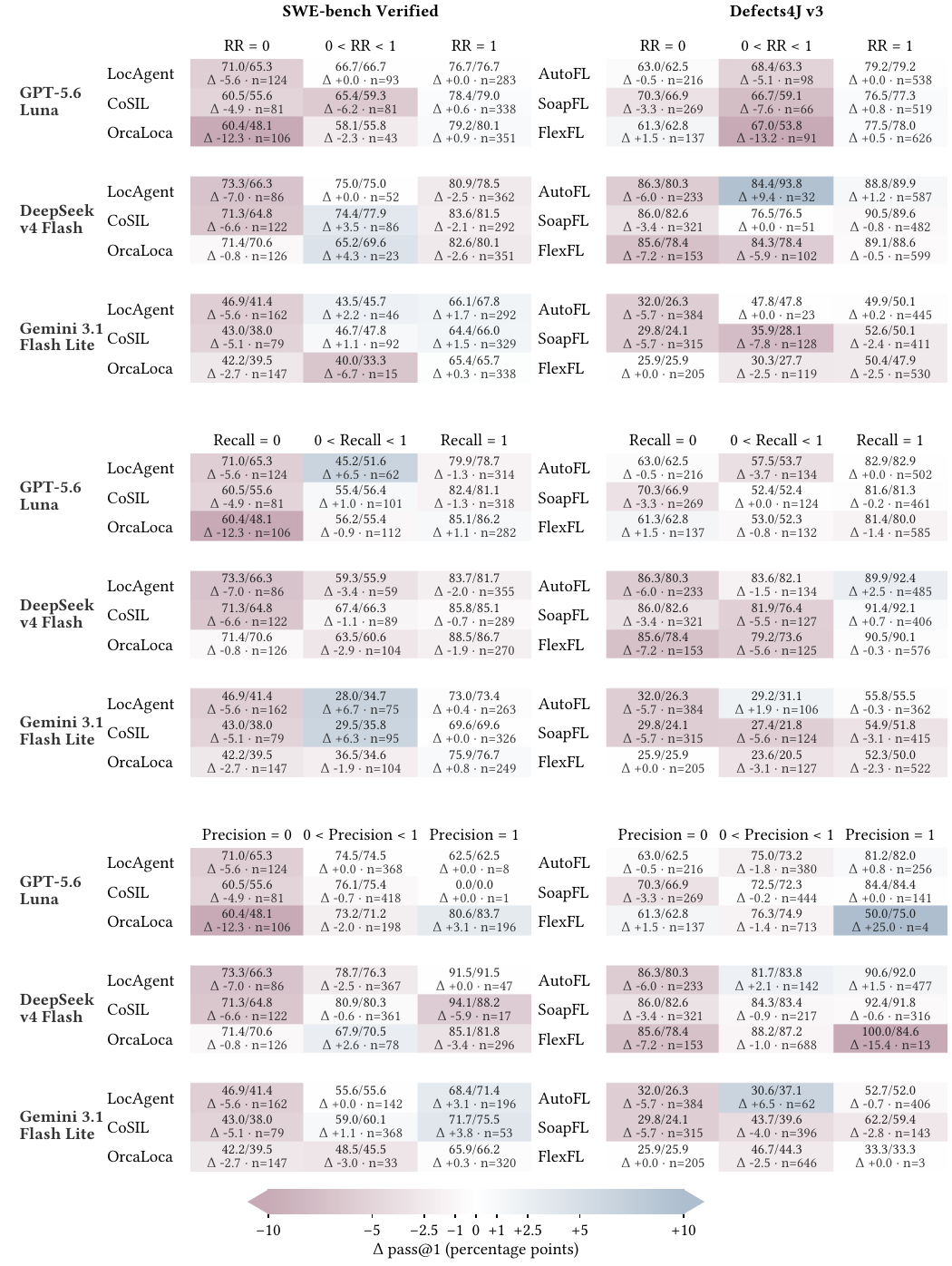}
\caption{Paired repair effects by RR, recall, and precision across the 18 With-FL settings. Each cell shows No-FL/With-FL pass@1 (\%), the paired change in percentage points, and the group size.}
\Description{Heatmaps of paired repair effects in three groups for each localization-quality metric, grouped by benchmark, repair model, and FL method.}
\label{fig:rq1-2-quality_heatmaps}
\end{figure}

\subsubsection{RQ1.2: How does repair benefit vary with localization quality?}
\label{sec:RQ1-utility}
\leavevmode\par


\phead{Design.}
RQ1.1 shows that FL changes the repair outcome for individual bugs but does not consistently improve overall repair success.
We next examine whether these effects vary with localization quality.
For each bug in the 18 With-FL settings, we measure the quality of its FL output using reciprocal rank (RR), recall, and precision, and group bugs by localization quality.
Within each group, we compare the repair outcomes with and without FL.
We evaluate each FL output against the gold entities used in that setting.
RR measures how highly the first correct entity is ranked, recall measures how much of the gold set is covered, and precision measures how much of the predicted set is correct.
Let $P$ be the set of predicted entities and $G$ the gold set.

\begin{itemize}[leftmargin=*]
\item \textbf{Reciprocal rank (RR).}
RR captures how highly the FL output ranks a correct location.
It is $1/r$ when the first correct entity appears at rank $r$, and zero if no correct entity is retrieved.
We divide bugs into three groups: miss ($RR=0$), later hit ($0<RR<1$), and Top-1 hit ($RR=1$).

\item \textbf{Recall.}
Recall measures how much of the faulty locations are covered by the FL output.
For a nonempty gold set, recall is $|P \cap G| / |G|$.
We divide bugs into miss (recall $=0$), partial ($0<$ recall $<1$), and complete (recall $=1$).

\item \textbf{Precision.}
Precision measures how much of the FL output points to correct locations rather than irrelevant ones.
For a nonempty prediction set, precision is $|P \cap G| / |P|$.
We assign precision zero to empty predictions.
We divide bugs into none correct (precision $=0$), mixed ($0<$ precision $<1$), and all correct (precision $=1$).
\end{itemize}

\findings{Higher localization quality does not guarantee additional repair benefit.}
Figure~\ref{fig:rq1-2-quality_heatmaps} shows how the change in pass@1 varies across localization-quality groups.
The highest-quality groups do not consistently show larger repair gains.
For RR, the change in pass@1 for Top-1 hits ranges from \(-2.56\) to \(+1.71\) percentage points, while the change for complete recall ranges from \(-3.13\) to \(+2.47\) points.
Even when the first predicted entity is correct or all gold entities are retrieved, the effect on repair success remains small and varies in direction.
The all-correct precision groups contain few bugs in several settings, making comparisons with other precision groups less reliable.

\results{Bugs with high-quality localization are often also easier to repair without FL.}
As shown in Figure~\ref{fig:rq1-2-quality_heatmaps}, the No-FL repair rate is generally higher in the strongest localization groups.
Across all 18 With-FL settings, the No-FL pass@1 is higher for Top-1 hits ($RR=1$) than for misses ($RR=0$).
For example, on Defects4J, GPT-5.6 Luna with AutoFL has a No-FL pass@1 of 79.2\% for Top-1 hits ($N=538$), compared with 63.0\% for misses ($N=216$).
On SWE-bench Verified, Gemini 3.1 Flash Lite with LocAgent shows the same pattern, with 66.1\% for Top-1 hits ($N=292$) versus 46.9\% for misses ($N=162$).
Similarly, the complete-recall group has a higher No-FL pass@1 than the miss group in all 18 With-FL settings.
This suggests that bugs that are easier to localize are often also easier for the agent to repair without FL.

\results{In contrast, low-quality localization more often reduces repair success.}
The largest decreases in the RR groups include GPT-5.6 Luna with FlexFL in the later-hit group (\(-13.19\) points), GPT-5.6 Luna with OrcaLoca in the miss group (\(-12.26\) points), Gemini 3.1 Flash Lite with SoapFL in the later-hit group (\(-7.81\) points), and GPT-5.6 Luna with SoapFL in the later-hit group (\(-7.58\) points).
These decreases are larger than any improvement observed in the Top-1 or complete-recall groups.
Together, these results suggest that the benefit of FL depends not only on localization quality, but also on how easily the bug can already be repaired without localization guidance.

\rqanswer{1}{The studied FL methods offer little benefit to repair success. Targets with high-quality FL are often already easier to repair without guidance, while inaccurate locations can introduce noise into the repair process.}

\subsection{RQ2: How does FL affect the agent's repair process?}
\label{sec:RQ2}

While RQ1 examines repair success, RQ2 examines how FL affects the repair process. In particular, we analyze 16,248 distinct repair trajectories from the 18 With-FL settings and six No-FL baselines in RQ1, including 6,000 from SWE-bench Verified and 10,248 from Defects4J.
We also manually analyze 160 distinct trajectories, sampling 40 each from No-FL and three With-FL groups defined by their match with gold locations: Correct FL contains only correct locations, Partial FL contains both correct and incorrect locations, and Wrong FL contains no correct locations.

\begin{table*}[tp]
\centering
\caption{Trajectory behavior across the 18 With-FL settings and their corresponding No-FL baselines.}
\label{tab:rq2-trajectory-summary}
\footnotesize
\setlength{\tabcolsep}{1.5pt}
\renewcommand{\arraystretch}{1.06}
\textit{(a) SWE-bench Verified}\par\smallskip
\begin{tabular*}{\textwidth}{@{\extracolsep{\fill}}ll *{3}{W{c}{35pt}} W{c}{38pt} W{c}{35pt} W{r}{13pt} *{2}{W{r}{35pt}}@{}}
\toprule
\multirow{2}{*}{Model} & \multirow{2}{*}{FL method} & \multicolumn{3}{c}{Navigation} & \multicolumn{2}{c}{FL adoption (\%)} & \multicolumn{3}{c}{Misleading FL} \\
\cmidrule(lr){3-5}\cmidrule(lr){6-7}\cmidrule(l){8-10}
& & \makecell{Reach\\(\%)} & \makecell{Reach\\step} & \makecell{Pre-edit\\files} & \makecell{Initial\\inspection} & \makecell{First\\edit} & $N$ & \makecell{Anchored\\$a$ (\%)} & \makecell{Recovered\\$r$ (\%)} \\
\midrule
\multirow{3}{*}{\makecell[l]{GPT-5.6\\Luna}} & LocAgent & 99.0$\to$98.8 & 1.60$\to$1.09 & 2.98$\to$3.24 & 95.2$\to$100.0 & 95.8$\to$96.0 & 21 & 21 (100.0) & 14 (66.7) \\
 & CoSIL & 99.0$\to$99.0 & 1.60$\to$1.14 & 2.98$\to$3.67 & 95.0$\to$99.8 & 94.2$\to$94.6 & 27 & 26 (96.3) & 23 (88.5) \\
 & OrcaLoca & 99.0$\to$98.0 & 1.60$\to$1.18 & 2.98$\to$2.53 & 89.4$\to$99.8 & 92.6$\to$95.0 & 40 & 39 (97.5) & 29 (74.4) \\
\addlinespace[4pt]
\multirow{3}{*}{\makecell[l]{DeepSeek-v4\\Flash}} & LocAgent & 99.4$\to$99.4 & 2.40$\to$1.23 & 3.74$\to$2.80 & 93.6$\to$99.8 & 62.5$\to$60.9 & 7 & 7 (100.0) & 4 (57.1) \\
 & CoSIL & 99.4$\to$99.2 & 2.40$\to$1.90 & 3.74$\to$3.33 & 87.6$\to$96.4 & 55.7$\to$50.1 & 54 & 40 (74.1) & 38 (95.0) \\
 & OrcaLoca & 99.4$\to$99.6 & 2.40$\to$1.54 & 3.74$\to$2.90 & 85.2$\to$99.2 & 58.1$\to$58.7 & 52 & 49 (94.2) & 47 (95.9) \\
\addlinespace[4pt]
\multirow{3}{*}{\makecell[l]{Gemini 3.1\\Flash Lite}} & LocAgent & 96.6$\to$95.0 & 4.93$\to$2.89 & 1.98$\to$1.78 & 77.6$\to$91.2 & 75.2$\to$78.4 & 40 & 33 (82.5) & 14 (42.4) \\
 & CoSIL & 96.6$\to$96.4 & 4.93$\to$4.16 & 1.98$\to$1.68 & 85.2$\to$86.0 & 77.3$\to$78.3 & 23 & 18 (78.3) & 9 (50.0) \\
 & OrcaLoca & 96.6$\to$95.2 & 4.93$\to$3.02 & 1.98$\to$1.64 & 70.6$\to$89.8 & 70.4$\to$75.5 & 70 & 57 (81.4) & 38 (66.7) \\
\bottomrule
\end{tabular*}
\par\medskip
\textit{(b) Defects4J v3.0.1}\par\smallskip
\begin{tabular*}{\textwidth}{@{\extracolsep{\fill}}ll *{3}{W{c}{35pt}} W{c}{38pt} W{c}{35pt} W{r}{13pt} *{2}{W{r}{35pt}}@{}}
\toprule
\multirow{2}{*}{Model} & \multirow{2}{*}{FL method} & \multicolumn{3}{c}{Navigation} & \multicolumn{2}{c}{FL adoption (\%)} & \multicolumn{3}{c}{Misleading FL} \\
\cmidrule(lr){3-5}\cmidrule(lr){6-7}\cmidrule(l){8-10}
& & \makecell{Reach\\(\%)} & \makecell{Reach\\step} & \makecell{Pre-edit\\files} & \makecell{Initial\\inspection} & \makecell{First\\edit} & $N$ & \makecell{Anchored\\$a$ (\%)} & \makecell{Recovered\\$r$ (\%)} \\
\midrule
\multirow{3}{*}{\makecell[l]{GPT-5.6\\Luna}} & AutoFL & 98.4$\to$98.8 & 1.64$\to$1.40 & 4.14$\to$3.73 & 81.2$\to$94.0 & 84.8$\to$86.3 & 139 & 92 (66.2) & 84 (91.3) \\
 & SoapFL & 98.4$\to$97.3 & 1.64$\to$1.53 & 4.14$\to$3.72 & 78.4$\to$99.6 & 77.8$\to$84.9 & 193 & 191 (99.0) & 170 (89.0) \\
 & FlexFL & 98.4$\to$98.2 & 1.64$\to$1.30 & 4.14$\to$3.75 & 86.6$\to$96.7 & 89.8$\to$92.4 & 87 & 63 (72.4) & 52 (82.5) \\
\addlinespace[4pt]
\multirow{3}{*}{\makecell[l]{DeepSeek-v4\\Flash}} & AutoFL & 99.2$\to$98.7 & 3.61$\to$2.56 & 4.99$\to$4.52 & 68.8$\to$86.8 & 71.5$\to$76.0 & 168 & 58 (34.5) & 51 (87.9) \\
 & SoapFL & 99.2$\to$98.7 & 3.61$\to$2.97 & 4.99$\to$4.57 & 75.0$\to$98.9 & 67.9$\to$72.2 & 175 & 171 (97.7) & 162 (94.7) \\
 & FlexFL & 99.2$\to$99.2 & 3.61$\to$2.41 & 4.99$\to$4.54 & 88.3$\to$97.6 & 81.1$\to$82.7 & 87 & 65 (74.7) & 61 (93.8) \\
\addlinespace[4pt]
\multirow{3}{*}{\makecell[l]{Gemini 3.1\\Flash Lite}} & AutoFL & 88.6$\to$87.9 & 3.68$\to$1.76 & 2.15$\to$1.93 & 70.7$\to$98.5 & 81.1$\to$91.6 & 172 & 158 (91.9) & 55 (34.8) \\
 & SoapFL & 88.6$\to$87.1 & 3.68$\to$2.36 & 2.15$\to$2.00 & 71.7$\to$99.6 & 76.9$\to$88.3 & 206 & 203 (98.5) & 99 (48.8) \\
 & FlexFL & 88.6$\to$92.3 & 3.68$\to$1.94 & 2.15$\to$2.32 & 85.9$\to$98.9 & 88.7$\to$95.4 & 104 & 95 (91.3) & 38 (40.0) \\
\bottomrule
\end{tabular*}
\end{table*}

\subsubsection{RQ2.1: How does FL affect the agent's repository navigation?}
\label{sec:RQ2-navigation}
\leavevmode\par

\phead{Design.}
We examine whether and how quickly the repair agent reaches a gold file.
We consider the agent to have reached the right location when it inspects a file modified in the gold patch.
For No-FL and With-FL trajectories, we measure three aspects of navigation: whether the agent reaches a gold file, the first step at which it does so, and the number of distinct production files inspected before the first source edit.
A step is one model response, including all tool calls issued in that response.
Gold-file reach is measured over all 854 targets in each Defects4J setting and all 500 targets in each SWE-bench Verified setting.
The first-reach step is computed only for trajectories that reach a gold file in the corresponding setting.
All FL methods for a given benchmark and model share the same No-FL navigation baseline.

\findings{FL mainly helps agents reach the right location earlier rather than making them more likely to reach it.}
As shown in the Navigation columns of Table~\ref{tab:rq2-trajectory-summary}, No-FL gold-file reach across the 18 paired comparisons ranges from 88.6\% to 99.4\%, compared with 87.1\% to 99.6\% With-FL.
Reach rates are already high without FL and change only slightly with guidance.
Among trajectories that do reach a gold file, however, the mean first-reach step is lower With-FL in all 18 settings, by 0.11--2.04 steps.
The mean number of distinct production files inspected before the first source edit is also lower With-FL in 15 settings, by 0.15--0.94 files, and higher in three settings, by 0.17--0.69 files.
These differences compare the displayed means rather than paired changes for the same targets.
The main navigation benefit of FL is therefore a shorter path to relevant code.


The manual study highlights how FL changes agents' starting points for navigation.
In the 40 No-FL trajectories, agents identify candidate code from bug reports or failing-test information: 30 search for provided symbols, seven directly open files using explicit location clues, and three infer production class names from test names before searching.
Further inspection of tests, call relationships, or implementations occurs in 38/40 trajectories and leads agents to revise their initial localization judgment in 13/40.
With FL, agents initially target recommended files or symbols in 102/120 trajectories: 48/102 search for symbols or filenames, 41/102 directly access specified files.
Thus, No-FL agents derive navigation entry points from the original bug information, while With-FL agents frequently start from recommended candidates.

\subsubsection{RQ2.2: How does FL influence the agent's search and edit decisions?}
\label{sec:RQ2-adoption}
\leavevmode\par

\phead{Design.}
We measure whether the agent follows FL guidance during initial code inspection and when making edits.
We consider the guidance adopted if the corresponding model response involves at least one FL-recommended file.
We compare paired With-FL and No-FL runs for the same target, retaining only pairs with an assessable FL scope and the corresponding observable response in both settings.
On SWE-bench Verified, initial-inspection rates use 500 pairs per setting, and first-edit rates use 497--500 pairs.
On Defects4J, these denominators range from 792 to 852 and from 753 to 851, respectively.
For the No-FL run, we use the same FL-recommended files only as a reference to see whether the agent would have visited them without receiving the guidance.
The difference in adoption rates shows how much FL changes the agent's search and edit decisions.

\findings{Agents consistently follow FL during initial navigation on both benchmarks.}
As shown in the FL Adoption columns of Table~\ref{tab:rq2-trajectory-summary}, FL increases initial-inspection adoption in all 18 With-FL settings.
The increases range from 0.8 to 27.9 percentage points, and With-FL initial-inspection adoption ranges from 86.0\% to 100.0\%.
These results suggest that agents use FL-recommended files as starting points, even when they would otherwise begin their search elsewhere.

\results{FL has less influence on where agents first edit on SWE-bench Verified, but continues to influence their edit decisions on Defects4J.}
On SWE-bench Verified, first-edit adoption increases in seven of the nine settings and decreases in two, with changes ranging from \(-5.6\) to \(+5.1\) percentage points.
In eight of the nine settings, the absolute change is smaller for first edit than for initial inspection.
The exception is Gemini 3.1 Flash Lite with CoSIL, where first-edit adoption increases by 1.0 percentage point and initial-inspection adoption by 0.8 percentage points.
On Defects4J, first-edit adoption increases in all nine settings, by 1.5--11.4 percentage points.
Although these increases are smaller than those for initial inspection, they show that FL guidance continues to influence edit decisions in this benchmark.
The different first-edit patterns suggest that the influence of FL beyond initial navigation varies across repair contexts.

The manual study shows how agents further assess the recommended locations before editing.
With Correct FL, 29/40 trajectories inspect code before editing, and seven revise their explanation of the fault after inspection.
With Partial FL, 30/40 compare multiple candidates to identify the responsible location, while another eight continue investigating the cause from a selected entry point.
For example, with Partial FL on Math-63, the agent follows the array-comparison code to the helper that performs the comparison.
These cases show how recommended locations can serve as starting points while subsequent inspection informs the edit.

\subsubsection{RQ2.3: How does the agent respond to misleading FL guidance?}
\label{sec:RQ2-misleading}
\leavevmode\par

\phead{Design.}
We examine how agents respond to completely misleading FL guidance.
We define an FL result as completely misleading when none of the files exposed to the repair agent overlaps a canonical gold file.
For these cases, we measure how often agents initially follow the guidance and later reach a gold file.
A trajectory is initially anchored when its first production-source inspection includes at least one FL-recommended file.
Among initially anchored trajectories, we define recovery as inspecting or editing a gold file in a later step.
We report the anchoring rate over all completely misleading cases and the recovery rate over initially anchored trajectories.

\findings{Agents often follow misleading FL at first, but many later reach a gold file.}
As shown in the Misleading FL columns of Table~\ref{tab:rq2-trajectory-summary}, all 18 With-FL settings contain completely misleading FL cases, ranging from 7 to 206 targets.
Among these cases, initial anchoring ranges from 34.5\% to 100.0\%, showing that agents frequently begin their search from locations suggested by FL even when none of those locations is correct.
Recovery among initially anchored trajectories ranges from 34.8\% to 95.9\%.
More than half of anchored trajectories later reach a gold file in 13 of the 18 With-FL settings, although recovery remains below 50\% in all three Gemini 3.1 Flash Lite Defects4J settings.
Thus, incorrect FL can redirect the agent's early search, but subsequent exploration often allows the agent to move beyond that initial guidance.

The manual study illustrates how this recovery occurs.
Among the Wrong FL cases, 29 have completely misleading file-level guidance and initially follow a recommended file, matching the criteria above.
Of these, 16 later reach a gold file.
Observed ways of moving beyond the recommendations include following call relationships, ruling out a suggested explanation using contrary evidence, and searching directly from issue clues.
For example, in Mockito-15, the agent follows the call chain to the location that performs the assignment.
Across all 40 Wrong FL trajectories, 23 retain their main edit in the recommended implementation and 17 move elsewhere, with nine and 13 passing the official evaluation, respectively.
These observations suggest that agents can move beyond incorrect FL guidance and still produce successful repairs.

\subsubsection{RQ2.4: How does FL affect repair effort and end-to-end cost?}
\label{sec:RQ2-effort}
\leavevmode\par

\phead{Design.}
We examine how FL affects repair effort using agent steps and token usage across the 18 With-FL settings. We first compare the resources used during repair with and without FL, including the number of steps and the total input and output tokens. We then measure resource use for the full pipeline by adding the tokens consumed by FL to the repair tokens. This allows us to examine whether FL reduces repair effort and whether these savings remain after including the cost of localization.

\begin{table*}[t]
\centering
\caption{Average per-target resource use across repair models.}
\label{tab:rq2-cost}
\footnotesize
\setlength{\tabcolsep}{2pt}
\renewcommand{\arraystretch}{1.0}
\textit{(a) SWE-bench Verified}\par\smallskip
\begin{tabular*}{\textwidth}{@{\extracolsep{\fill}}W{l}{40pt} *{9}{W{r}{32pt}}@{}}
\toprule
\multirow{2}{*}{Setting} & \multicolumn{3}{c}{GPT-5.6 Luna} & \multicolumn{3}{c}{DeepSeek-v4 Flash} & \multicolumn{3}{c}{Gemini 3.1 Flash Lite} \\
\cmidrule(lr){2-4}\cmidrule(lr){5-7}\cmidrule(l){8-10}
& \makecell{Repair\\tokens} & \makecell{Repair\\steps} & \makecell{Total\\tokens} & \makecell{Repair\\tokens} & \makecell{Repair\\steps} & \makecell{Total\\tokens} & \makecell{Repair\\tokens} & \makecell{Repair\\steps} & \makecell{Total\\tokens} \\
\midrule
No-FL & 138.7K & 12.2 & 138.7K & 1.50M & 43.6 & 1.50M & 573.9K & 47.3 & 573.9K \\
\addlinespace[3pt]
LocAgent & 123.3K & 11.5 & 183.1K & 1.03M & 33.3 & 1.26M & 402.1K & 39.7 & 819.4K \\
 & ($-11.1$\%) & ($-6.0$\%) & ($+32.0$\%) & ($-31.3$\%) & ($-23.5$\%) & ($-16.4$\%) & ($-29.9$\%) & ($-16.1$\%) & ($+42.8$\%) \\
\addlinespace[3pt]
CoSIL & 138.1K & 12.0 & 279.0K & 1.29M & 39.1 & 1.75M & 500.6K & 45.3 & 692.7K \\
 & ($-0.4$\%) & ($-1.7$\%) & ($+101.2$\%) & ($-13.9$\%) & ($-10.3$\%) & ($+16.6$\%) & ($-12.8$\%) & ($-4.2$\%) & ($+20.7$\%) \\
\addlinespace[3pt]
OrcaLoca & 129.6K & 12.0 & 318.7K & 1.19M & 37.7 & 1.31M & 484.7K & 43.3 & 595.6K \\
 & ($-6.5$\%) & ($-2.1$\%) & ($+129.8$\%) & ($-21.0$\%) & ($-13.4$\%) & ($-12.9$\%) & ($-15.5$\%) & ($-8.4$\%) & ($+3.8$\%) \\
\bottomrule
\end{tabular*}
\par\medskip
\textit{(b) Defects4J v3.0.1}\par\smallskip
\begin{tabular*}{\textwidth}{@{\extracolsep{\fill}}W{l}{40pt} *{9}{W{r}{32pt}}@{}}
\toprule
\multirow{2}{*}{Setting} & \multicolumn{3}{c}{GPT-5.6 Luna} & \multicolumn{3}{c}{DeepSeek-v4 Flash} & \multicolumn{3}{c}{Gemini 3.1 Flash Lite} \\
\cmidrule(lr){2-4}\cmidrule(lr){5-7}\cmidrule(l){8-10}
& \makecell{Repair\\tokens} & \makecell{Repair\\steps} & \makecell{Total\\tokens} & \makecell{Repair\\tokens} & \makecell{Repair\\steps} & \makecell{Total\\tokens} & \makecell{Repair\\tokens} & \makecell{Repair\\steps} & \makecell{Total\\tokens} \\
\midrule
No-FL & 129.4K & 9.5 & 129.4K & 802.0K & 25.3 & 802.0K & 292.4K & 26.8 & 292.4K \\
\addlinespace[3pt]
AutoFL & 110.7K & 9.1 & 130.2K & 880.6K & 25.0 & 936.2K & 252.3K & 23.9 & 287.6K \\
 & ($-14.5$\%) & ($-4.4$\%) & ($+0.6$\%) & ($+9.8$\%) & ($-1.2$\%) & ($+16.7$\%) & ($-13.7$\%) & ($-10.7$\%) & ($-1.7$\%) \\
\addlinespace[3pt]
SoapFL & 111.9K & 9.1 & 139.0K & 839.5K & 25.3 & 899.3K & 261.7K & 24.0 & 286.0K \\
 & ($-13.6$\%) & ($-3.7$\%) & ($+7.4$\%) & ($+4.7$\%) & ($+0.2$\%) & ($+12.1$\%) & ($-10.5$\%) & ($-10.4$\%) & ($-2.2$\%) \\
\addlinespace[3pt]
FlexFL & 108.7K & 8.9 & 145.3K & 775.0K & 24.0 & 795.2K & 274.6K & 24.3 & 294.9K \\
 & ($-16.0$\%) & ($-6.0$\%) & ($+12.2$\%) & ($-3.4$\%) & ($-5.0$\%) & ($-0.8$\%) & ($-6.1$\%) & ($-9.3$\%) & ($+0.9$\%) \\
\bottomrule
\end{tabular*}
\end{table*}

\findings{FL usually reduces downstream repair effort, but localization overhead often outweighs these savings end to end.}
Table~\ref{tab:rq2-cost} compares repair-stage and end-to-end resource use.
Within the repair stage, token use decreases in 16 of the 18 With-FL settings, with changes ranging from a 31.3\% decrease to a 9.8\% increase.
Steps decrease in 17 settings, with changes ranging from a 23.5\% decrease to a 0.2\% increase.
These reductions are consistent with the shorter navigation paths observed in RQ2.1.
Once localization itself is included, however, end-to-end token use increases in 13 settings and decreases in only five, with changes ranging from a 16.4\% decrease to a 129.8\% increase.
All six GPT-5.6 Luna settings increase end-to-end token use.
The five decreases occur for DeepSeek-v4 Flash with FlexFL, LocAgent, and OrcaLoca, and for Gemini 3.1 Flash Lite with AutoFL and SoapFL.
Overall, localization often consumes more tokens than it saves during repair.


\rqanswer{2}{FL helps agents reach relevant code sooner and shapes their early search. Agents can also localize faults on their own, check suggested locations, and adjust their search as they learn more. Location information can reduce agent token use, but the studied FL methods often consume more tokens than they save overall.}

\subsection{RQ3: When does FL provide additional value for agentic repair?}
\label{sec:RQ3}


RQ1 and RQ2 examine how current FL methods affect repair success and the repair process.
Most FL research focuses on issue descriptions and failing-test information.
Yet RQ2 shows that agents use clues in the bug input to guide their search, suggesting that the value of FL may vary with the repair context.
We therefore extend our study to contexts that differ in input information and repair complexity.
We remove direct location clues from SWE-bench Verified issues and include two additional benchmarks.
SWE-bench Multimodal provides both text and images, while the SWE-bench Pro bug subset involves larger and more distributed bugs.
We use gold locations and bug-cause descriptions derived from gold patches to study the potential value of different forms of FL guidance independently of the accuracy of existing methods.
Finally, we examine whether conventional FL ground truth captures the locations used by successful repairs.



\subsubsection{RQ3.1: Is FL more useful when bug inputs provide fewer location clues?}
\label{sec:RQ3-issue-input}
\leavevmode\par

\phead{Design.}
We construct localization-reduced issues (LR-issues) for all 500 SWE-bench Verified tasks using mini-swe-agent with GPT-5.6 Luna.
Given the original issue, project code, and gold patch, the agent removes direct location clues while preserving the reported behavior and repair requirements.
We include bug-cause descriptions alongside gold locations to study different forms of FL output.
We generate these descriptions using mini-swe-agent with the bug input, pre-fix source code, and gold patch.
The prompt asks for the cause of the failure without repair locations or fix instructions.
For both original issues and LR-issues, we compare No-FL with different combinations of gold location and bug-cause information using pass@1, repair tokens, and agent steps.

\begin{table}[t]
\centering
\caption{Repair success and resource use with GPT-5.6 Luna across three SWE-bench benchmarks.}
\label{tab:rq3-issue-input}
\footnotesize
\setlength{\tabcolsep}{4pt}
\renewcommand{\arraystretch}{1.12}
\newcommand{\metricchange}[2]{#1~(#2)}
\begin{tabular*}{\linewidth}{@{\extracolsep{\fill}}llrrr@{}}
\toprule
\textbf{Issue input} & \textbf{FL input} & \textbf{pass@1} & \textbf{Repair tokens} & \textbf{Steps} \\
\midrule
\multicolumn{5}{@{}l}{\textbf{SWE-bench Verified} ($N=500$)} \\
\addlinespace[3pt]
Issue & No-FL & 73.4\% & 138.7K & 12.24 \\
 & Gold file & \metricchange{74.2\%}{+0.8 pp} & \metricchange{114.6K}{$-17.4$\%} & \metricchange{11.21}{$-8.4$\%} \\
 & Gold symbol & \metricchange{76.2\%}{+2.8 pp} & \metricchange{101.7K}{$-26.7$\%} & \metricchange{10.85}{$-11.4$\%} \\
 & Bug cause & \metricchange{77.8\%}{+4.4 pp} & \metricchange{105.0K}{$-24.3$\%} & \metricchange{10.83}{$-11.5$\%} \\
 & Gold symbol + bug cause & \metricchange{80.0\%}{+6.6 pp} & \metricchange{92.8K}{$-33.1$\%} & \metricchange{10.49}{$-14.3$\%} \\
\addlinespace[5pt]
LR-issue & No-FL & 68.4\% & 186.9K & 12.91 \\
 & Gold file & \metricchange{72.4\%}{+4.0 pp} & \metricchange{119.8K}{$-35.9$\%} & \metricchange{11.59}{$-10.2$\%} \\
 & Gold symbol & \metricchange{73.0\%}{+4.6 pp} & \metricchange{113.7K}{$-39.2$\%} & \metricchange{11.46}{$-11.2$\%} \\
 & Bug cause & \metricchange{75.4\%}{+7.0 pp} & \metricchange{135.1K}{$-27.7$\%} & \metricchange{11.58}{$-10.3$\%} \\
 & Gold symbol + bug cause & \metricchange{81.4\%}{+13.0 pp} & \metricchange{97.4K}{$-47.9$\%} & \metricchange{10.88}{$-15.7$\%} \\
\midrule
\multicolumn{5}{@{}l}{\textbf{SWE-bench Multimodal} ($N=480$)} \\
\addlinespace[3pt]
Issue + images & No-FL & 33.3\% & 231.9K & 12.86 \\
 & Gold file & \metricchange{34.6\%}{+1.3 pp} & \metricchange{193.0K}{$-16.8$\%} & \metricchange{12.26}{$-4.6$\%} \\
 & Gold symbol & \metricchange{35.2\%}{+1.9 pp} & \metricchange{188.9K}{$-18.5$\%} & \metricchange{12.29}{$-4.4$\%} \\
 & Bug cause & \metricchange{40.0\%}{+6.7 pp} & \metricchange{197.0K}{$-15.0$\%} & \metricchange{12.13}{$-5.7$\%} \\
 & Gold symbol + bug cause & \metricchange{42.1\%}{+8.8 pp} & \metricchange{168.9K}{$-27.2$\%} & \metricchange{11.63}{$-9.6$\%} \\
\midrule
\multicolumn{5}{@{}l}{\textbf{SWE-bench Pro} (bug subset, $N=221$)} \\
\addlinespace[3pt]
Issue & No-FL & 50.2\% & 211.3K & 12.87 \\
 & Gold file & \metricchange{52.0\%}{+1.8 pp} & \metricchange{187.4K}{$-11.3$\%} & \metricchange{12.05}{$-6.4$\%} \\
 & Gold symbol & \metricchange{53.8\%}{+3.6 pp} & \metricchange{181.5K}{$-14.1$\%} & \metricchange{11.87}{$-7.8$\%} \\
 & Bug cause & \metricchange{52.9\%}{+2.7 pp} & \metricchange{203.2K}{$-3.8$\%} & \metricchange{12.41}{$-3.6$\%} \\
 & Gold symbol + bug cause & \metricchange{52.9\%}{+2.7 pp} & \metricchange{181.0K}{$-14.3$\%} & \metricchange{11.86}{$-7.8$\%} \\
\bottomrule
\end{tabular*}
\end{table}

\findings{FL yields larger gains in repair success when issues contain fewer direct location clues.}
On SWE-bench Verified, No-FL pass@1 falls from 73.4\% with the original issues to 68.4\% with LR-issues.
The gain over No-FL from gold files increases from 0.8 to 4.0 percentage points, while the gain from gold symbols increases from 2.8 to 4.6 points.
The same pattern holds for bug causes, whose gain increases from 4.4 to 7.0 points, and for gold symbols with bug causes, whose gain increases from 6.6 to 13.0 points.
Repair-token reductions also increase from 17.4\% to 35.9\% for gold files and from 26.7\% to 39.2\% for gold symbols.
Even when bug causes are already provided, adding gold symbols improves pass@1 by 6.0 points with LR-issues, compared with 2.2 points with the original issues.
These results suggest that the additional value of FL depends on the location clues already available in the bug input, supporting its use on demand rather than as a mandatory preprocessing step for every repair.

\subsubsection{RQ3.2: How do different forms of FL output affect repair across settings?}
\label{sec:RQ3-output}
\leavevmode\par

\phead{Design.}
We include two additional benchmarks to study FL across different repair contexts.
SWE-bench Multimodal~\cite{yang2024swebenchmultimodalaisystems} includes 480 tasks with both text and images as bug input.
The SWE-bench Pro bug subset~\cite{scale2025swebenchpro} includes 221 tasks with issue descriptions and larger, more distributed repairs.
Patches modify 3.85 files on average, and 82.8\% span multiple files, compared with 1.25 files and 14.2\% in SWE-bench Verified.
Across these benchmarks and the original and reduced issues in Verified, we compare No-FL with different combinations of gold location and bug-cause information using pass@1, repair tokens, and agent steps.

\findings{Gold locations reduce repair tokens in every setting, but improve repair success only modestly.}
In Table~\ref{tab:rq3-issue-input}, providing gold files or gold symbols improves pass@1 by 0.8--4.6 percentage points over No-FL.
Thus, even accurate locations bring only modest gains in repair success.
They also reduce mean repair tokens by 11.3--39.2\% across the four settings.
Gold symbols use fewer repair tokens than gold files in every setting, suggesting that more precise locations can further reduce token use.

\results{Bug causes can help repair even when locations are given, but the gains vary across settings.}
Bug causes alone outperform gold symbols in pass@1 by 1.6 percentage points on SWE-bench Verified with original issues, 2.4 points with LR-issues, and 4.8 points on SWE-bench Multimodal.
On the SWE-bench Pro bug subset, however, they achieve 52.9\% pass@1, below the 53.8\% achieved with gold symbols.
Combining bug causes with gold symbols gives the highest observed pass@1 on Verified with both issue inputs and on Multimodal, improving over gold symbols alone by 3.8, 8.4, and 6.9 points, respectively.
On Pro, the combination remains at 52.9\%, showing that the benefit of adding bug causes varies across settings.
In terms of cost, bug causes alone reduce repair tokens less than gold symbols in all four settings.
On Pro, the reductions relative to No-FL are 3.8\% and 14.1\%, respectively.
Adding bug causes to gold symbols further reduces token use on Verified and Multimodal, but barely changes it on Pro, from 181.5K to 181.0K.

\begin{table}[tbp]
\centering
\caption{Method-level scope, line-level scope, and patch diversity of SWE-bench Verified and Defects4J v3.0.1.}
\label{tab:rq3-granularity}
\footnotesize
\setlength{\tabcolsep}{6pt}
\renewcommand{\arraystretch}{1.14}
\begin{tabularx}{\linewidth}{@{}Xrr@{}}
\toprule
\textbf{Measure} & \textbf{SWE-bench Verified} & \textbf{Defects4J v3.0.1} \\
\midrule
\multicolumn{3}{@{}l}{\textit{(a) Method-level scope}} \\
Zero-method tasks & 41/500 (8.2\%) & 27/854 (3.2\%) \\
\quad Module/class scope & 19 & 16 \\
\quad New method & 18 & 9 \\
\quad Mixed/non-code & 4 & 2 \\
\midrule
\multicolumn{3}{@{}l}{\textit{(b) Line-level scope}} \\
Only insertion & 100/500 (20.0\%) & 246/854 (28.8\%) \\
At least one insertion-only location & 262/500 (52.4\%) & 543/854 (63.6\%) \\
\midrule
\multicolumn{3}{@{}l}{\textit{(c) Patch diversity among successful repairs}} \\
Cases with $\geq 2$ successful patches & 1,095 & 1,851 \\
Modified files differ & 127 (11.6\%) & 184 (9.9\%) \\
Modified lines differ & 430 (39.3\%) & 827 (44.7\%) \\
\bottomrule
\end{tabularx}
\end{table}

\subsubsection{RQ3.3: Does conventional FL ground truth adequately represent valid repair locations?}
\label{sec:RQ3-granularity}
\leavevmode\par

\phead{Design.}
FL ground truth is typically drawn from benchmark reference patches.
We examine two possible limitations of this practice on SWE-bench Verified and Defects4J.
First, we check whether repairs can be represented by existing methods or lines.
Method-level localization is widely used and has been reported as preferred by developers~\cite{kochhar2016practitioners}.
We use Pylance for Python and the Eclipse JDT Language Server for Java to identify the symbols containing each change.
We manually classify patches with no changes inside existing methods into module/class-level changes, new methods, and mixed or non-code changes.
We also count patches with only insertions and those with at least one insertion-only location, since added lines do not exist in the buggy code.
Second, we check whether successful repairs for the same bug modify different locations.
For each model and target, we compare patches when at least two repair settings succeed.
We record whether these patches modify the same files and the same lines in the buggy code.

\findings{Conventional FL ground truth does not fully capture valid repair locations in repository-level repair.}
Existing methods and lines cannot represent all repair locations.
Table~\ref{tab:rq3-granularity}(a) shows that all changes fall outside existing method bodies in 3.2\% of Defects4J v3.0.1 tasks and 8.2\% of SWE-bench Verified tasks.
At the line level, pure-insertion patches account for 28.8\% and 20.0\% of tasks, respectively, while 63.6\% and 52.4\% contain at least one insertion-only location (Table~\ref{tab:rq3-granularity}(b)).
These insertions have no corresponding line in the buggy code, so targets restricted to existing lines cannot fully represent them.
A single reference patch also misses differences among successful repairs.
Among model--target pairs with at least two successful patches, modified lines differ in 44.7\% of Defects4J v3.0.1 cases and 39.3\% of SWE-bench Verified cases.
Modified files differ in 9.9\% and 11.6\% of cases, respectively (Table~\ref{tab:rq3-granularity}(c)).
Thus, ground truth drawn from one reference patch does not capture all locations used by successful repairs.

\rqanswer{3}{FL has greater potential when bug inputs provide fewer location clues.
Accurate locations reduce repair tokens but only modestly improve repair success, while bug-cause information can further increase the success rate.
Conventional FL ground truth does not fully capture the locations used by successful repairs.}

\section{Implications}
\label{Discussion}

\phead{Move toward on-demand localization and make efficiency a priority.}
Our results suggest that FL may not be needed before every repair, since agents can often locate relevant code on their own and inaccurate guidance can mislead their initial search.
This motivates studying when to invoke FL, such as when the bug input provides few location clues or the agent struggles to find relevant code.
The cost of providing this guidance also needs attention, as localization often consumes more tokens than it saves during repair.
Future methods should therefore consider efficiency alongside accuracy.

\phead{Study FL across diverse repair contexts and consider information beyond code locations.}
RQ3 suggests that the value of FL depends on both the repair context and the information it provides.
Future studies should cover a broader range of settings, including inputs with fewer location clues, multimodal inputs, and larger, more distributed repairs, and examine what guidance helps in each setting.
Our results show that bug-cause information can help beyond accurate locations, although the benefit varies across settings.
There are also limits to how locations are represented, since some repairs fall outside existing methods or lines and successful patches may modify different locations.
Together, these findings motivate studying how FL can complement locations with diagnostic information, such as failure behavior or evidence explaining why a location is suspicious.

\phead{Reconsider the formulation of FL beyond location ranking and benchmark FL in downstream repair contexts.}
Our findings suggest that ranking suspicious code locations may not fully capture the role of FL in agentic repair.
Higher location accuracy does not necessarily lead to better repair outcomes.
Reference patches also represent particular repair choices and may not cover all locations useful for understanding or fixing a bug.
Furthermore, the potential benefits of bug-cause information suggest that agents may need guidance on why a failure occurs as well as where to look.
These observations motivate exploring FL formulations that combine candidate locations with explanations of the failure and supporting evidence.
Evaluating such guidance requires considering both its accuracy and its usefulness in the repair process.
Future research should therefore examine what information FL should provide and how to assess its contribution to understanding and repairing faults.

\section{Threats to Validity}
\label{Threats}

\phead{Internal validity.}
Our manual analysis covers 160 trajectories and may involve subjective judgments or miss less common behaviors.
We interpret model-generated explanations together with observable tool actions, as underlying reasoning traces are not consistently accessible.
LLM randomness may also affect our results, as we run each task once per setting given the scale of our experiments and the high cost of repeated runs.
Our coverage of 44 settings across different models, FL methods, and repair contexts helps reduce reliance on individual runs when drawing overall conclusions.
In RQ3, issue rewriting may unintentionally alter information beyond location clues, although the prompt requires preserving the reported behavior and repair requirements.

\phead{External validity.}
We provide FL guidance before repair, following the common setup of existing FL methods.
Our findings may differ for on-demand or iterative FL.
We select SWE-bench Verified and Defects4J for reproducibility and coverage of common FL inputs, but both may overlap with model training data.
Paired comparisons use the same model and task in both conditions, although prior exposure may still affect the benefit of FL.
RQ3 also includes the SWE-bench Pro bug subset, whose source repositories use strong copyleft licenses to reduce contamination risk~\cite{scale2025swebenchpro}, without guaranteeing its absence.

\section{Conclusion}
\label{sec:conclusion}

We studied the role of FL in agentic repair.
We evaluated six recent FL methods on two benchmarks with three repair models under a common agentic repair framework.
The studied FL methods do not consistently improve repair success, although they guide early search and often reduce repair effort.
Accurate locations substantially reduce repair token use, but the overhead of current FL methods often outweighs downstream token savings.
Experiments with gold guidance suggest greater potential when bug inputs contain fewer location clues, and show that bug-cause information can help beyond accurate locations.
We also find that conventional FL ground truth does not fully capture the locations used by successful repairs.
Together, these findings motivate studying on-demand localization, providing diagnostic information alongside locations, and evaluating FL through its contribution to repair success and end-to-end efficiency.


\bibliographystyle{ACM-Reference-Format}
\bibliography{references}

@inproceedings{jones2005empirical,
  title={Empirical evaluation of the tarantula automatic fault-localization technique},
  author={Jones, James A and Harrold, Mary Jean},
  booktitle={Proceedings of the 20th IEEE/ACM international Conference on Automated software engineering},
  pages={273--282},
  year={2005}
}

@inproceedings{zhou2012should,
  title={Where should the bugs be fixed? more accurate information retrieval-based bug localization based on bug reports},
  author={Zhou, Jian and Zhang, Hongyu and Lo, David},
  booktitle={2012 34th International conference on software engineering (ICSE)},
  pages={14--24},
  year={2012},
  organization={IEEE}
}

@inproceedings{chen2025locagent,
  title={Locagent: Graph-guided llm agents for code localization},
  author={Chen, Zhaoling and Tang, Robert and Deng, Gangda and Wu, Fang and Wu, Jialong and Jiang, Zhiwei and Prasanna, Viktor and Cohan, Arman and Wang, Xingyao},
  booktitle={Proceedings of the 63rd Annual Meeting of the Association for Computational Linguistics (Volume 1: Long Papers)},
  pages={8697--8727},
  year={2025}
}

@inproceedings{jiang2025issue,
  title={Issue localization via llm-driven iterative code graph searching},
  author={Jiang, Zhonghao and Ren, Xiaoxue and Yan, Meng and Jiang, Wei and Li, Yong and Liu, Zhongxin},
  booktitle={2025 40th IEEE/ACM International Conference on Automated Software Engineering (ASE)},
  pages={3034--3045},
  year={2025},
  organization={IEEE}
}

@inproceedings{kochhar2016practitioners,
  title={Practitioners' expectations on automated fault localization},
  author={Kochhar, Pavneet Singh and Xia, Xin and Lo, David and Li, Shanping},
  booktitle={Proceedings of the 25th international symposium on software testing and analysis},
  pages={165--176},
  year={2016}
}

@article{kang2024quantitative,
  title={A quantitative and qualitative evaluation of LLM-based explainable fault localization},
  author={Kang, Sungmin and An, Gabin and Yoo, Shin},
  journal={Proceedings of the ACM on Software Engineering},
  volume={1},
  number={FSE},
  pages={1424--1446},
  year={2024},
  publisher={ACM New York, NY, USA}
}

@ARTICLE{10891926,
  author={Qin, Yihao and Wang, Shangwen and Lou, Yiling and Dong, Jinhao and Wang, Kaixin and Li, Xiaoling and Mao, Xiaoguang},
  journal={IEEE Transactions on Software Engineering}, 
  title={SoapFL: A Standard Operating Procedure for LLM-Based Method-Level Fault Localization}, 
  year={2025},
  volume={51},
  number={4},
  pages={1173-1187},
  doi={10.1109/TSE.2025.3543187}}

@inproceedings{yao2023react,
  title = {{ReAct}: Synergizing Reasoning and Acting in Language Models},
  author = {Yao, Shunyu and Zhao, Jeffrey and Yu, Dian and Du, Nan and Shafran, Izhak and Narasimhan, Karthik and Cao, Yuan},
  booktitle = {International Conference on Learning Representations (ICLR) },
  year = {2023},
  html = {https://arxiv.org/abs/2210.03629},
}

@inproceedings{wang2025openhands,
  title={Openhands: An open platform for ai software developers as generalist agents},
  author={Wang, Xingyao and Li, Boxuan and Song, Yufan and Xu, Frank F and Tang, Xiangru and Zhuge, Mingchen and Pan, Jiayi and Song, Yueqi and Li, Bowen and Singh, Jaskirat and others},
  booktitle={International Conference on Learning Representations},
  volume={2025},
  pages={65882--65919},
  year={2025}
}

@inproceedings{tang2024codeagent,
  title={Codeagent: Autonomous communicative agents for code review},
  author={Tang, Xunzhu and Kim, Kisub and Song, Yewei and Lothritz, Cedric and Li, Bei and Ezzini, Saad and Tian, Haoye and Klein, Jacques and Bissyand{\'e}, Tegawend{\'e} F},
  booktitle={Proceedings of the 2024 Conference on Empirical Methods in Natural Language Processing},
  pages={11279--11313},
  year={2024}
}

@article{yang2024swe,
  title={Swe-agent: Agent-computer interfaces enable automated software engineering},
  author={Yang, John and Jimenez, Carlos and Wettig, Alexander and Lieret, Kilian and Yao, Shunyu and Narasimhan, Karthik and Press, Ofir},
  journal={Advances in Neural Information Processing Systems},
  volume={37},
  pages={50528--50652},
  year={2024}
}

@inproceedings{bouzenia2025repairagent,
  title={Repairagent: An autonomous, llm-based agent for program repair},
  author={Bouzenia, Islem and Devanbu, Premkumar and Pradel, Michael},
  booktitle={2025 IEEE/ACM 47th International Conference on Software Engineering (ICSE)},
  pages={2188--2200},
  year={2025},
  organization={IEEE}
}

@inproceedings{sohn2017fluccs,
  title={Fluccs: Using code and change metrics to improve fault localization},
  author={Sohn, Jeongju and Yoo, Shin},
  booktitle={Proceedings of the 26th ACM SIGSOFT international symposium on software testing and analysis},
  pages={273--283},
  year={2017}
}

@inproceedings{lou2021boosting,
  title={Boosting coverage-based fault localization via graph-based representation learning},
  author={Lou, Yiling and Zhu, Qihao and Dong, Jinhao and Li, Xia and Sun, Zeyu and Hao, Dan and Zhang, Lu and Zhang, Lingming},
  booktitle={Proceedings of the 29th ACM joint meeting on european software engineering conference and symposium on the foundations of software engineering},
  pages={664--676},
  year={2021}
}

@inproceedings{abreu2007accuracy,
  title={On the accuracy of spectrum-based fault localization},
  author={Abreu, Rui and Zoeteweij, Peter and Van Gemund, Arjan JC},
  booktitle={Testing: Academic and industrial conference practice and research techniques-MUTATION (TAICPART-MUTATION 2007)},
  pages={89--98},
  year={2007},
  organization={IEEE}
}

@inproceedings{li2019deepfl,
  title={Deepfl: Integrating multiple fault diagnosis dimensions for deep fault localization},
  author={Li, Xia and Li, Wei and Zhang, Yuqun and Zhang, Lingming},
  booktitle={Proceedings of the 28th ACM SIGSOFT international symposium on software testing and analysis},
  pages={169--180},
  year={2019}
}

@inproceedings{bohme2017bug,
  title={Where is the bug and how is it fixed? an experiment with practitioners},
  author={B{\"o}hme, Marcel and Soremekun, Ezekiel O and Chattopadhyay, Sudipta and Ugherughe, Emamurho and Zeller, Andreas},
  booktitle={Proceedings of the 2017 11th joint meeting on foundations of software engineering},
  pages={117--128},
  year={2017}
}

@article{alaboudi2021exploratory,
  title={An exploratory study of debugging episodes},
  author={Alaboudi, Abdulaziz and LaToza, Thomas D},
  journal={arXiv preprint arXiv:2105.02162},
  year={2021}
}

@article{chen2021demystifying,
  title={Demystifying the challenges and benefits of analyzing user-reported logs in bug reports},
  author={Chen, An Ran and Chen, Tse-Hsun and Wang, Shaowei},
  journal={Empirical Software Engineering},
  volume={26},
  number={1},
  pages={8},
  year={2021},
  publisher={Springer}
}

@article{shi2025enhancing,
  title={Enhancing LLM-based Fault Localization with a Functionality-Aware Retrieval-Augmented Generation Framework},
  author={Shi, Xinyu and Li, Zhenhao and Chen, An Ran},
  journal={arXiv preprint arXiv:2509.20552},
  year={2025}
}

@inproceedings{3780338.3783277,
author = {Yu, Zhongming and Zhang, Hejia and Zhao, Yujie and Huang, Hanxian and Yao, Matrix and Ding, Ke and Zhao, Jishen},
title = {OrcaLoca: an LLM agent framework for software issue localization},
year = {2025},
publisher = {JMLR.org},
booktitle = {Proceedings of the 42nd International Conference on Machine Learning},
articleno = {2939},
numpages = {21},
location = {Vancouver, Canada},
series = {ICML'25}
}

@article{liu2026graphlocator,
  title={Graphlocator: Graph-guided causal reasoning for issue localization},
  author={Liu, Wei and Peng, Chao and Gao, Pengfei and Liu, Aofan and Zhang, Wei and Zhao, Haiyan and Jin, Zhi},
  journal={Proceedings of the ACM on Software Engineering},
  volume={3},
  number={FSE},
  pages={1128--1149},
  year={2026},
  publisher={ACM New York, NY, USA}
}

@article{tamoyan2026sherloc,
  title={Sherloc: Structured diagnostic localization for code repair agents},
  author={Tamoyan, Hovhannes and Narenthiran, Sean and Arakelyan, Erik and Mezini, Mira and Ginsburg, Boris},
  journal={arXiv preprint arXiv:2606.24820},
  year={2026}
}

@inproceedings{gangi2026swerank,
  title={SWERank: Software Issue Localization with Code Ranking},
  author={Gangi Reddy, Revanth and Suresh, Tarun and Doo, JaeHyeok and Liu, Ye and Nguyen, Xuan-Phi and Zhou, Yingbo and Yavuz, Semih and Xiong, Caiming and Ji, Heng and Joty, Shafiq},
  booktitle={International Conference on Learning Representations},
  volume={2026},
  pages={78997--79011},
  year={2026}
}

@article{xu2025flexfl,
  title={Flexfl: Flexible and effective fault localization with open-source large language models},
  author={Xu, Chuyang and Liu, Zhongxin and Ren, Xiaoxue and Zhang, Gehao and Liang, Ming and Lo, David},
  journal={IEEE Transactions on Software Engineering},
  volume={51},
  number={5},
  pages={1455--1471},
  year={2025},
  publisher={IEEE}
}

@article{rafi2024multi,
  title={A multi-agent approach to fault localization via graph-based retrieval and reflexion},
  author={Rafi, Md Nakhla and Kim, Dong Jae and Chen, Tse-Hsun and Wang, Shaowei},
  journal={arXiv preprint arXiv:2409.13642},
  year={2024}
}

@article{yeo2025improving,
  title={Improving llm-based fault localization with external memory and project context},
  author={Yeo, Inseok and Ryu, Duksan and Baik, Jongmoon},
  journal={arXiv preprint arXiv:2506.03585},
  year={2025}
}

@inproceedings{weimer2009automatically,
  title={Automatically finding patches using genetic programming},
  author={Weimer, Westley and Nguyen, ThanhVu and Le Goues, Claire and Forrest, Stephanie},
  booktitle={2009 IEEE 31st International Conference on Software Engineering},
  pages={364--374},
  year={2009},
  organization={IEEE}
}

@inproceedings{nguyen2013semfix,
  title={Semfix: Program repair via semantic analysis},
  author={Nguyen, Hoang Duong Thien and Qi, Dawei and Roychoudhury, Abhik and Chandra, Satish},
  booktitle={2013 35th International Conference on Software Engineering (ICSE)},
  pages={772--781},
  year={2013},
  organization={IEEE}
}

@inproceedings{wei2023copiloting,
  title={Copiloting the copilots: Fusing large language models with completion engines for automated program repair},
  author={Wei, Yuxiang and Xia, Chunqiu Steven and Zhang, Lingming},
  booktitle={Proceedings of the 31st ACM Joint European Software Engineering Conference and Symposium on the Foundations of Software Engineering},
  pages={172--184},
  year={2023}
}

@inproceedings{xia2024automated,
  title={Automated program repair via conversation: Fixing 162 out of 337 bugs for \$0.42 each using chatgpt},
  author={Xia, Chunqiu Steven and Zhang, Lingming},
  booktitle={Proceedings of the 33rd ACM SIGSOFT international symposium on software testing and analysis},
  pages={819--831},
  year={2024}
}

@article{xia2024agentless,
  title={Agentless: Demystifying llm-based software engineering agents},
  author={Xia, Chunqiu Steven and Deng, Yinlin and Dunn, Soren and Zhang, Lingming},
  journal={arXiv preprint arXiv:2407.01489},
  year={2024}
}

@inproceedings{zhang2024autocoderover,
  title={Autocoderover: Autonomous program improvement},
  author={Zhang, Yuntong and Ruan, Haifeng and Fan, Zhiyu and Roychoudhury, Abhik},
  booktitle={Proceedings of the 33rd ACM SIGSOFT International Symposium on Software Testing and Analysis},
  pages={1592--1604},
  year={2024}
}

@misc{minisweagent2026docs,
  author = {{SWE-agent Team}},
  title = {{mini-SWE-agent} Documentation},
  year = {2026},
  url = {https://mini-swe-agent.com/v2/},
  urldate = {2026-09-18}
}

@article{sepidband2026rgfl,
  title={Rgfl: Reasoning guided fault localization for automated program repair using large language models},
  author={Sepidband, Melika and Taherkhani, Hamed and Pham, Hung Viet and Hemmati, Hadi},
  journal={arXiv preprint arXiv:2601.18044},
  year={2026}
}

@article{townsend2026study,
  title={A Study on the Impact of Fault localization Granularity for Repository-Scale Code Repair Tasks},
  author={Townsend, Joseph and Pravin, Chandresh and Ngan, Kwun Ho and Parizy, Matthieu},
  journal={arXiv preprint arXiv:2604.00167},
  year={2026}
}

@article{sepidband2026role,
  title={On the Role of Fault Localization Context for LLM-Based Program Repair},
  author={Sepidband, Melika and Pham, Hung Viet and Hemmati, Hadi},
  journal={arXiv preprint arXiv:2604.05481},
  year={2026}
}

@misc{awad2026loc2repairframeworkevaluatingimpact,
      title={Loc2Repair: A Framework for Evaluating the Impact of File-Level Issue Localization in Repo-Level LLM Repair}, 
      author={Mohammad Nour Al Awad and Sergey Ivanov},
      year={2026},
      eprint={2606.30963},
      archivePrefix={arXiv},
      primaryClass={cs.SE},
      url={https://arxiv.org/abs/2606.30963}, 
}

@misc{swebenchverifieddocs,
  author = {{SWE-bench}},
  title = {{SWE-bench} Datasets},
  year = {n.d.},
  url = {https://github.com/SWE-bench/SWE-bench/blob/main/docs/guides/datasets.md},
  urldate = {2026-09-26}
}

@misc{defects4jrepo,
  author = {{Defects4J}},
  title = {{Defects4J} GitHub Repository},
  year = {n.d.},
  url = {https://github.com/rjust/defects4j},
  urldate = {2026-09-26}
}

@misc{openai2026gpt56luna,
  author = {{OpenAI}},
  title = {{GPT-5.6 Luna}},
  year = {2026},
  url = {https://developers.openai.com/api/docs/models/gpt-5.6-luna},
  urldate = {2026-09-18}
}

@misc{deepseek2026v4flash,
  author = {{DeepSeek-AI}},
  title = {{DeepSeek-V4-Flash} Model Card},
  year = {2026},
  url = {https://huggingface.co/deepseek-ai/DeepSeek-V4-Flash},
  urldate = {2026-09-18}
}

@misc{deepseek2026apiupdates,
  author = {{DeepSeek}},
  title = {{DeepSeek API} Change Log},
  year = {2026},
  url = {https://api-docs.deepseek.com/updates/},
  urldate = {2026-09-18}
}

@misc{google2026gemini31flashlite,
  author = {{Google}},
  title = {{Gemini 3.1 Flash-Lite}},
  year = {2026},
  url = {https://ai.google.dev/gemini-api/docs/models/gemini-3.1-flash-lite},
  urldate = {2026-09-18}
}

@misc{yang2024swebenchmultimodalaisystems,
      title={SWE-bench Multimodal: Do AI Systems Generalize to Visual Software Domains?}, 
      author={John Yang and Carlos E. Jimenez and Alex L. Zhang and Kilian Lieret and Joyce Yang and Xindi Wu and Ori Press and Niklas Muennighoff and Gabriel Synnaeve and Karthik R. Narasimhan and Diyi Yang and Sida I. Wang and Ofir Press},
      year={2024},
      eprint={2410.03859},
      archivePrefix={arXiv},
      primaryClass={cs.CL},
      url={https://arxiv.org/abs/2410.03859}, 
}

@misc{scale2025swebenchpro,
  author = {{The Scale Research Team}},
  title = {{SWE-Bench Pro}: Raising the Bar for Agentic Coding},
  year = {2025},
  url = {https://scale.com/blog/swe-bench-pro},
  urldate = {2026-10-01}
}

@article{mcnemar1947sampling,
  author = {McNemar, Quinn},
  title = {Note on the Sampling Error of the Difference Between Correlated Proportions or Percentages},
  journal = {Psychometrika},
  year = {1947},
  volume = {12},
  number = {2},
  pages = {153--157},
  doi = {10.1007/BF02295996}
}

@article{holm1979simple,
  author = {Holm, Sture},
  title = {A Simple Sequentially Rejective Multiple Test Procedure},
  journal = {Scandinavian Journal of Statistics},
  year = {1979},
  volume = {6},
  number = {2},
  pages = {65--70},
  url = {https://www.jstor.org/stable/4615733}
}


\end{document}